\documentclass[
reprint,
superscriptaddress,
amsmath,
amssymb,
aps,
prb
]{revtex4-2}

\usepackage{graphicx}
\usepackage{dcolumn}
\usepackage[T1]{fontenc}
\usepackage[utf8]{inputenc}
\usepackage{siunitx}
\usepackage{xcolor}
\usepackage{threeparttable}
\usepackage{bm}
\usepackage{chngcntr}
\usepackage{hyperref}
\usepackage[normalem]{ulem}
\usepackage{booktabs}

\begin{document}

\preprint{APS/123-QED}

\title{Interface engineering of the anomalous Hall effect in Ni-based heterostructures}

\author{Mainak Ghosh}
\thanks{equal contribution}
 \affiliation{Department of Physics, Indian Institute of Technology Bombay,
Powai, Mumbai 400076, India}
\author{Kusampal Yadav}%
\thanks{equal contribution}

\affiliation{School of Physical Sciences, Indian Association for the Cultivation of Science, 2A and 2B Raja S.~C.~Mullick Road, Kolkata 700032, India}%

\author{Kalyan sarkar}%
\affiliation{School of Physical Sciences, Indian Association for the Cultivation of Science, 2A and 2B Raja S.~C.~Mullick Road, Kolkata 700032, India}%

 \author{Kousik Das}%
\affiliation{School of Physical Sciences, Indian Association for the Cultivation of Science, 2A and 2B Raja S.~C.~Mullick Road, Kolkata 700032, India}%

\author{Devajyoti Mukherjee}%
\email{sspdm@iacs.res.in}
\affiliation{School of Physical Sciences, Indian Association for the Cultivation of Science, 2A and 2B Raja S.~C.~Mullick Road, Kolkata 700032, India}%

\author{Sayantika Bhowal}
\email{sbhowal@iitb.ac.in}
 \affiliation{Department of Physics, Indian Institute of Technology Bombay,
Powai, Mumbai 400076, India}

\begin{abstract}
Using a combined experimental and first-principles theoretical approach, we demonstrate interface engineering of the anomalous Hall effect in Ni-based epitaxial thin-film heterostructures. Ferromagnetic Ni thin films are grown on (001)-oriented single-crystal LaAlO$_3$, SrTiO$_3$, and MgO substrates, which impose different biaxial tensile strains of 0.3\%, 0.6\%, and 0.8\%, respectively. Our room-temperature Hall transport measurements reveal a pronounced substrate-dependent modulation of the anomalous Hall conductivity. Interestingly, our calculations show that strain alone cannot account for the experimentally observed trends. Instead, we identify interfacial inversion-symmetry breaking, which induces Rashba spin-orbit interaction, as the key mechanism governing the anomalous Hall conductivity across different interfaces. Building on this understanding, we further demonstrate both theoretically and experimentally that the anomalous Hall conductivity can be continuously tuned by an external electric field. These findings establish the critical role of substrate-induced interfacial effects in controlling the anomalous Hall effect in engineered heterostructures and provide a viable pathway toward electrically tunable room-temperature spintronic devices.

\end{abstract}

\maketitle

\section{INTRODUCTION}

Emerging quantum materials, with their intriguing physical properties ranging from strong correlations to topological phases, are at the forefront of condensed matter physics research \cite{i1, i2, i3, i4, i7, i8, i9}. In particular, low-dimensional systems constitute a prominent class, as reduced dimensionality enhances quantum effects, leading to functionalities beyond those observed in conventional bulk materials \cite{il1, il2, il3}. Low-dimensional quantum materials are particularly attractive due to the high degree of tunability of their physical properties, such as magnetism, electronic structure, and spin–orbit coupling (SOC), through variations in structural parameters, external fields, chemical composition (e.g., doping or alloying), and strain, among other factors \cite{il5, il6, il7, il2, i14, i16}.

In this context, two-dimensional (2D) and quasi-2D materials are of particular interest, as quantum confinement and enhanced surface or interface effects significantly modify their physical properties compared to their bulk counterparts \cite{Das04072014, wines2025toward}. Despite their promise, many conventional 2D materials suffer from intrinsic limitations. For example, in most 2D magnets, long-range magnetic order persists only at low temperatures \cite{i11}, and the absence of strong interfacial fields limits the tunability of SOC and correlation effects \cite{i12}. In this context, quasi-2D interface heterostructures offer a promising route to overcome these limitations \cite{i4, i14, i15, i16}. Recent advances in atomic layer-controlled growth techniques have facilitated the fabrication of atomically sharp interfaces between different oxide materials \cite{mckee1998crystalline}. At such interfaces, the intrinsic properties of individual materials become strongly coupled through charge, lattice, spin, and orbital degrees of freedom. Consequently, symmetry breaking, structural coupling, and competing electrostatic and magnetic interactions can give rise to emergent quantum phenomena that are absent in the parent materials.

Heterostructures composed of transition metals and complex oxides provide a versatile platform for engineering novel electronic and spin-dependent phenomena at interfaces. In particular, interfaces between ferromagnetic metals and perovskite oxides have attracted considerable attention due to the interplay between magnetism, SOC, broken inversion symmetry, and charge reconstruction \cite{i17, i18, i19, i20, i21}. This interplay leads to a range of emergent phenomena, including modified magnetic anisotropy, interfacial exchange interactions, and spin-dependent transport responses \cite{Hua2023, Xiaomin2025, Prm031401, liu2012giant, ma2014interface}.

Such systems are central to modern spintronic and magnetoelectronic applications, including magnetic sensors, memory devices, and electrically tunable functionalities \cite{201502824, memresitivity, zhang20242d, dieny2017perpendicular}. The performance and stability of ferromagnetic metal/oxide interfaces are strongly influenced by the microscopic interfacial structure, including chemical bonding, interfacial reactions, morphological stability of the metallic layer, and the thermodynamic robustness of the oxide substrate. These factors are highly sensitive to growth conditions and operating temperatures \cite{zhou2024surface, dieny2017perpendicular}. Despite extensive experimental and theoretical efforts, achieving a comprehensive and predictive understanding of how interfacial chemistry and structural reconstructions govern the emergent properties of ferromagnetic metal/oxide heterostructures remains an open challenge \cite{MADEJ2025163379, zhou2024surface, 94PhysRevB, D0TC00311E}.

In this work, we address this challenge by growing epitaxial nickel (Ni) thin films on SrTiO$_3$ (STO), LaAlO$_3$ (LAO), and MgO substrates, investigating their anomalous Hall transport using a combined theoretical and experimental approach, and demonstrating a strategy to tune the anomalous Hall effect (AHE) at these interfaces. Our central finding is the identification of interfacial Rashba interaction as a key mechanism governing the AHE. Building on this insight, we demonstrate that the AHE can be effectively tuned at the interface by modulating the Rashba interaction through an external gate voltage. This tunability is established through systematic computational and experimental analysis of the voltage-dependent anomalous Hall conductivity (AHC) across Ni/STO, Ni/LAO, and Ni/MgO heterostructures.

Our findings highlight the crucial role of interface-driven symmetry breaking in controlling the AHC at heterostructures. In particular, we emphasize the significant effect of the substrate, extending beyond its conventional role in inducing interfacial strain. Beyond these fundamental insights, the voltage-controlled tuning of spin-polarized charge currents in the AHE, as demonstrated in our work, offers a promising route toward energy-efficient magnetization switching \cite{maruyama2009large, Bocirnea2020, 101063, Li2024}, with potential applications in next-generation low-power spintronic devices.

\section{RESULTS AND DISCUSSIONS}

We begin by discussing the epitaxial growth of Ni thin films on various oxide substrates, followed by our findings on the anomalous Hall effect in these heterostructures.

\subsection{Structural characterization}
\begin{figure*}[]
 \centering
    \includegraphics[width=1.8\columnwidth]{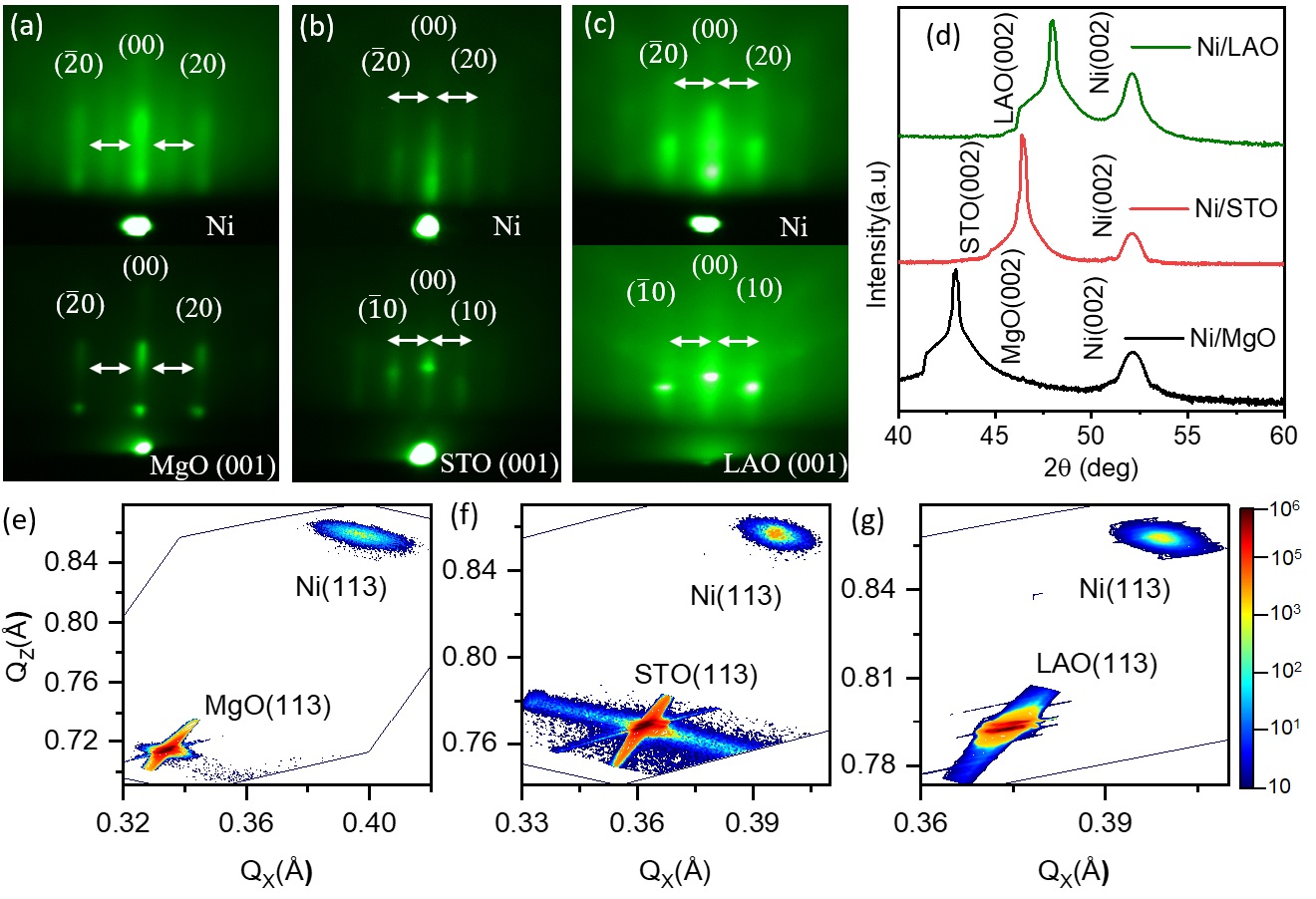}
    \caption{Structural characterization of epitaxial Ni thin films grown on oxide substrates. In-situ reflection high-energy electron diffraction (RHEED) patterns of (a) Ni/MgO, (b) Ni/STO, and (c) Ni/LAO heterostructures, respectively.  (d) X-ray diffraction (XRD) $\theta-2\theta$ patterns of Ni/MgO, Ni/STO, and Ni/LAO heterostructures. Reciprocal space maps acquired around the (113) reflections for (e) Ni/MgO, (f) Ni/STO, and (g) Ni/LAO heterostructures, respectively.}
    \label{fig_xrd}
\end{figure*}

Fig. \ref{fig_xrd} shows the high-quality epitaxial growth of Ni (bulk lattice parameter, $a_{\rm Ni}$ = 3.52~\AA,~JCPDS No. 00-004-0850) thin films deposited on single crystal MgO ($a_{\rm MgO}$ = 4.22~\AA), STO ($a_{\rm STO}$ = 3.91~\AA), and LAO ($a_{\rm LAO}$ = 3.79~\AA) substrates. Figs.\ref{fig_xrd}a-c show the in-situ reflection high-energy electron diffraction (RHEED) patterns of the Ni films and the corresponding bare substrates. The patterns exhibit sharp and continuous streaks, accompanied by pronounced Kikuchi lines, indicative of layer-by-layer growth, high crystalline quality, and predominantly two-dimensional epitaxial growth of (001)-oriented Ni films on all substrates \cite{Liu2020, Wu2024, Sampaio2025}. From the quantitative analysis of the RHEED patterns, we obtain the in-plane lattice parameters of the Ni films, viz., (3.56 $\pm$ 0.015), (3.55 $\pm$ 0.02), and (3.53 $\pm$ 0.03)~\AA~ for Ni/MgO, Ni/STO, and Ni/LAO substrates, respectively, indicating the presence of tensile strain in all films.

The epitaxial nature of the films is further confirmed by the $\theta$ – $2\theta$ X-ray diffraction (XRD) patterns, as shown in Fig. \ref{fig_xrd}d, for all heterostructures. Only the Ni (002) reflection, corresponding to the face-centered cubic (fcc) phase of Ni, is observed together with the (002) reflections of the single-crystalline MgO, STO, and LAO substrates, confirming the epitaxial growth of the constituent layers with a preferred (001) orientation. The quality of the epitaxial growth is further confirmed by rocking curve and azimuthal ($\Phi$) scan measurements, as discussed in the SM (see Fig. 1 of the SM \cite{supple_mat}).

The single-crystalline nature of the Ni heterostructures grown on MgO (001),  STO (001), and LAO (001) substrates is further validated by high-resolution XRD reciprocal space mapping (RSM) performed around the asymmetric (113) Bragg planes of the respective substrates \cite{Jiang2019}. Fig. \ref{fig_xrd}e-g display the RSMs for the Ni/MgO, Ni/STO, and Ni/LAO heterostructures, respectively. In all cases, a single, well-defined diffraction spot corresponding to the Ni (113) reflection of the fcc Ni phase is observed relative to the substrate peak, confirming the epitaxial growth and high crystalline quality of the Ni layers \cite{Yadav2025}. For the Ni/MgO heterostructure, the Ni (113) reflection is situated significantly away from the substrate peak, consistent with the large lattice mismatch ($\sim$17$\%$) between Ni and MgO. In contrast, the Ni reflections in the Ni/STO and Ni/LAO heterostructures are located closer to the corresponding substrate peaks, reflecting their comparatively smaller lattice mismatches of approximately 10$\%$ and 7$\%$, respectively.\\

From the RSM data, we extract the in-plane ($a$) and out-of-plane ($c$) lattice parameters of the Ni layers as well as the in-plane ($\epsilon_{\parallel}$) strain relative to bulk Ni. The obtained values are summarized in Table \ref{tab:table1}. The lattice parameters, obtained from the RSM analysis are consistent with the in-plane values derived from the RHEED measurements and with the out-of-plane parameters obtained from the $\theta$ – 2$\theta$ XRD scans. The values listed in Table \ref{tab:table1} indicate a substrate-dependent strain in the Ni layers, arising from the epitaxial growth on lattice-mismatched substrate.

\begin{table}[bt]
\centering
\normalsize
\setlength{\tabcolsep}{5pt}
\renewcommand{\arraystretch}{1.0}

\caption{\label{tab:table1}%
Out-of-plane ($c$) and in-plane ($a$) lattice parameters, corresponding to the parallel epitaxial strain ($\epsilon_{\parallel}$) for Ni films, grown on MgO, STO, and LAO substrates. $a_0$ is the bulk lattice parameter of Ni.}

\begin{threeparttable}

\begin{tabular}{lccc}
\toprule

Sample 
& $c$ (\AA) 
& $a$ (\AA) 
& $\epsilon_{\parallel}$ (\%) \\

& & & $\left(\frac{a}{a_{0}}\right)-1$ \\

\midrule

Ni/MgO & 3.49 ($\pm$0.03) & 3.55 ($\pm$0.03) & 0.8 \\
Ni/STO & 3.50 ($\pm$0.01) & 3.54 ($\pm$0.05) & 0.6 \\
Ni/LAO & 3.50 ($\pm$0.04) & 3.53 ($\pm$0.02) & 0.3 \\

\bottomrule

\end{tabular}

\end{threeparttable}
\end{table}

\subsection{Hall transport measurements}
\begin{figure}[h]
 \centering
    \includegraphics[width=\columnwidth]{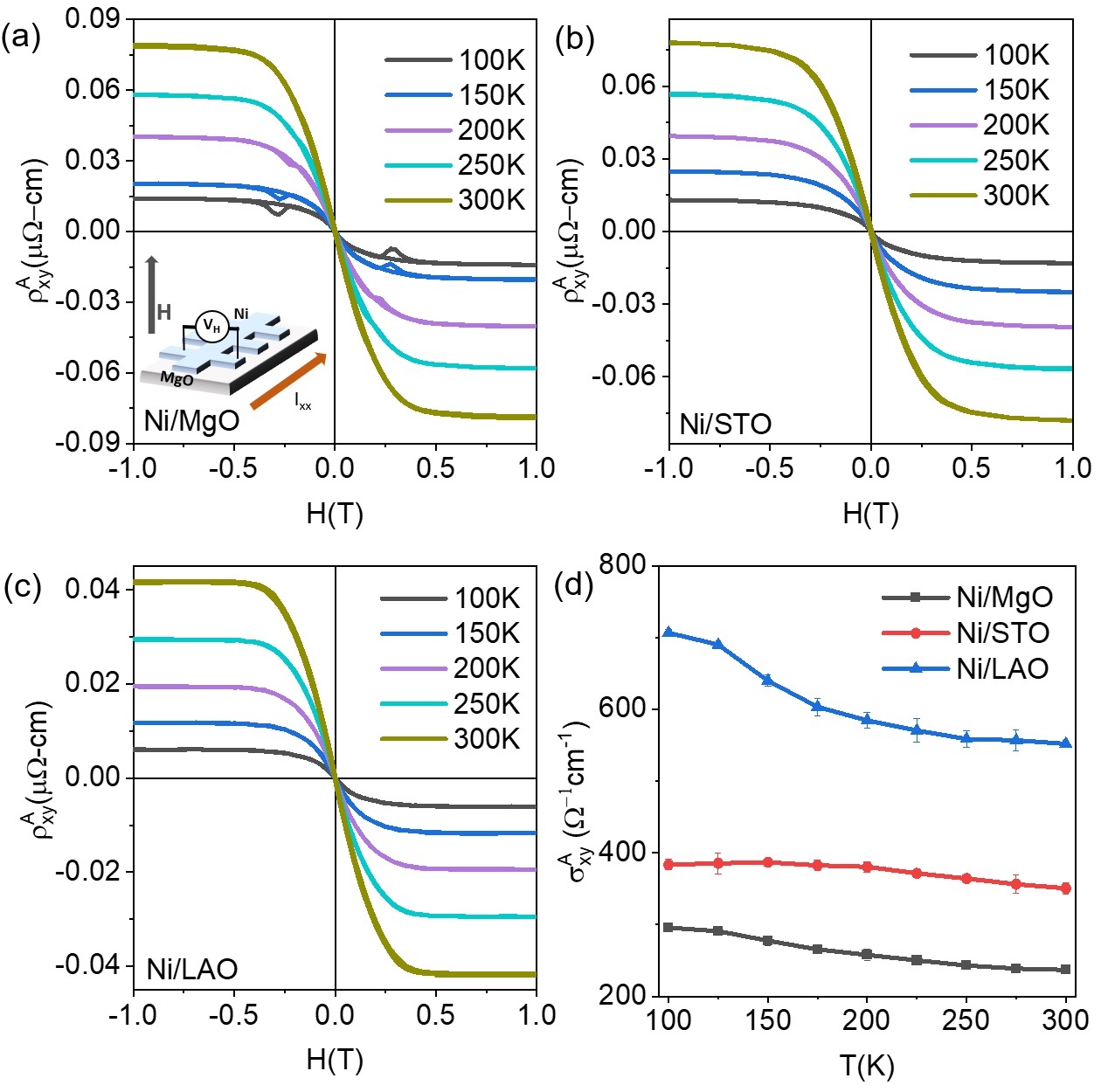}
    \caption{Temperature-dependent Hall properties of heterostructures, measured between 100 and 300 K. Field-dependent anomalous Hall resistivity for (a) Ni/MgO, (b) Ni/STO, and (c) Ni/LAO heterostructures, respectively. Inset shows the schematic for the Hall measurement configuration. (d) Temperature dependence of the maximum AHC $\sigma_{xy}^{A}$  for all heterostructures at 1 T magnetic field.}
    \label{fig_AHE}
\end{figure}

Next, we focus on the Hall transport measurements. The magnetic-field dependence of the Hall resistivity ($\rho_{xy}$), measured over a temperature range of 100–300 K, for the Ni/MgO, Ni/STO, and Ni/LAO heterostructures, using standard Hall bar devices, is shown in Fig. \ref{fig_AHE}a-c. The total Hall resistivity in a ferromagnetic thin film can be written as \cite{Venkateswara2023,Matsuno2016}, 
\begin{equation}
   \rho_{xy}(H)= \rho_{xy}^{O}+\rho_{xy}^{A},
    \label{eqTHE}
\end{equation}
where $\rho_{xy}^{O}$, $\rho_{xy}^{A}$ are respectively the ordinary Hall effect (OHE) and anomalous Hall effect (AHE) contributions to the Hall resistivity. We extract the carrier concentration for all heterostructures from the OHE coefficient through linear fitting, $\rho_{xy}^{O}$ = R$_O$H (see SM~\cite{supple_mat} section 2). We find electron-type charge carriers with similar carrier densities, viz., $1.3(\pm0.03)\times10^{23} $ cm$^{-3}$, $1.4(\pm0.01)\times10^{23} $ cm$^{-3}$, and $1.8(\pm0.03)\times10^{23} $ cm$^{-3}$ for Ni/MgO, Ni/STO, and Ni/LAO thin films at room temperature, respectively \cite{SINGH2024, MT1998} . As evident from Fig.~\ref{fig_AHE}a-c, $\rho_{xy}^{A}$ increases rapidly at low magnetic fields, exhibiting anomalous features up to $\sim$ 0.5 T for Ni/MgO and Ni/STO, and up to $\sim$0.4 T for Ni/LAO. After reaching saturation, the curves become linear, extending to 5 T (for clarity, we show only data up to 1T), consistent with the respective out-of-plane saturation fields (see section 3  of SM~\cite{supple_mat} \cite{MONDAL2022, Soya2025}).  Over the entire temperature range, the magnitude of $\rho_{xy}^{A}$ is largest for Ni/MgO, smaller for Ni/STO, and smallest for Ni/LAO, indicating a clear substrate-dependent modification of the anomalous Hall response. 
\\

We further obtain the intrinsic anomalous Hall conductivity (AHC) from the fitting of our measured data (see section 4 of SM~\cite{supple_mat} \cite{Nagaosa2010, Li2012,Zhang_2017,Lavine1961,Taewon2021,Kim2026}). Our obtained values are $\sim 180.2$ $    \Omega^{-1}\text{cm}^{-1}$, $\sim 343.7$ $ \Omega^{-1}\text{cm}^{-1}$, and $\sim 579.5$ $ \Omega^{-1}\text{cm}^{-1}$ for Ni/MgO, Ni/LAO, and Ni/STO, respectively. The intrinsic AHC coefficient attains its highest value for Ni/LAO and decreases progressively for Ni/STO and Ni/MgO, indicating a clear substrate-dependent evolution of the intrinsic Hall response. We calculate the AHC  at different temperatures using $\sigma_{xy} = -\frac{\rho_{xy}}{\rho_{xy}^2 + \rho_{xx}^2}$~\cite{Venkateswara2023} (see section 5 of the SM~\cite{supple_mat} for $\rho_{xx}$ data \cite{Wolowiec2022,Kumbhakar2025,Chen2009,Sokolov2016}). Fig. ~\ref{fig_AHE}d shows the temperature dependence of the maximum value of AHC ($\sigma^A_{xy}$) for Ni/MgO, Ni/LAO, and Ni/STO heterostructures. We note that the conductivity has a weak temperature dependence, suggesting the dominance of an intrinsic mechanism in governing the AHC~\cite{Cao2026,Wang2016, Bhattacharya2024, Bera2023}.

\subsection{Biaxial Strain vs. Interfacial Rashba Interaction}
To understand the underlying mechanisms behind the observed changes in the AHE across different heterostructures, we investigate these systems computationally using density functional theory (DFT). These results provide crucial insights into disentangling the roles of biaxial strain and interfacial Rashba interactions in governing the observed AHE behavior.

\subsubsection{Effects of Biaxial Strain on Bulk Ni}

As listed in Table \ref{tab:table1}, depending on the choice of substrate, Ni thin films undergo different biaxial tensile strains. To understand if the observed transport properties in the heterostructures simply result from these interfacial strains, we have carried out electronic structure calculations of bulk face-centered cubic (fcc) Ni in the presence of in-plane biaxial strain of up to $\pm 2\%$, where the $\pm$ sign refers to tensile and compressive strain, respectively. These calculations capture the strain-induced changes in the electronic structure and the resulting AHC.  

\begin{figure}[t]
 \includegraphics[width=\columnwidth]{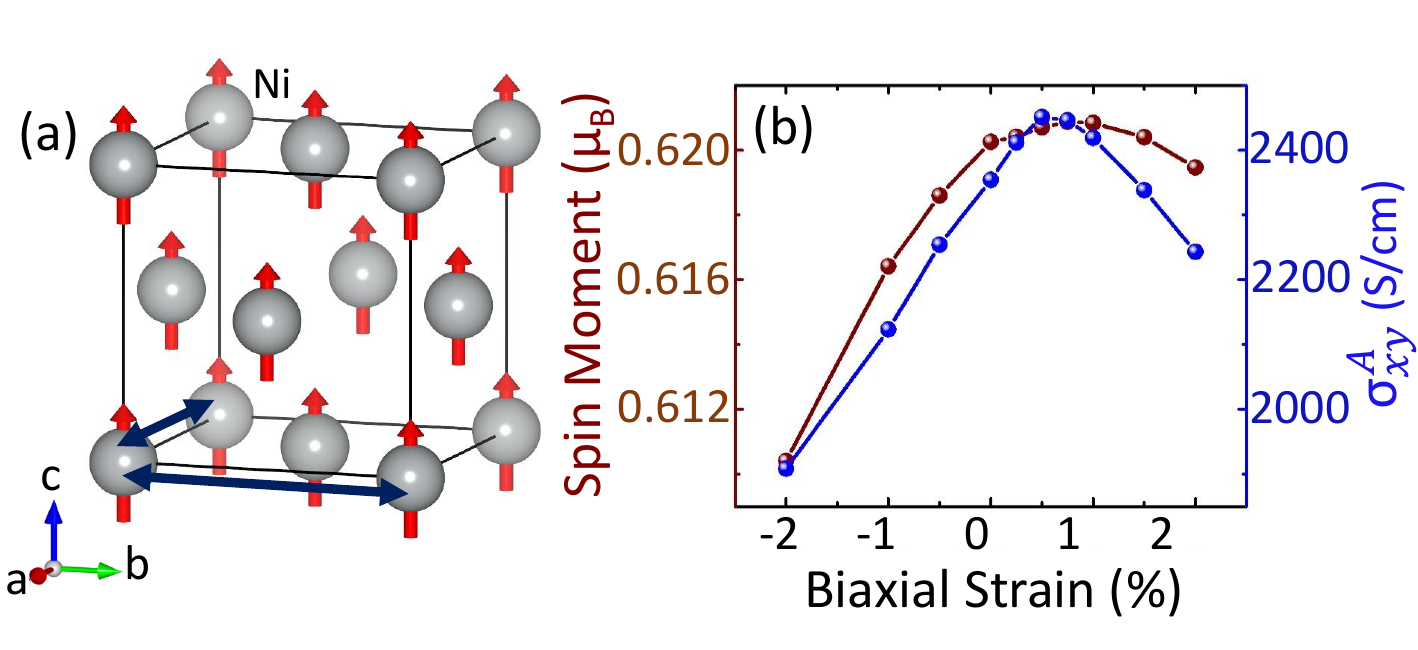}
  \caption{Effect of strain on bulk Ni. (a) Bulk fcc Ni crystal structure with spin magnetic moment along the $z$-axis. Here, the two-sided arrow indicates the direction of biaxial strain. (b) The variation of the total spin magnetic moment and AHC with biaxial strain in bulk Ni.}
 \label{fig1}
 \end{figure}

We note that the application of biaxial strain lowers the symmetry of fcc Ni from the cubic $Fm-3m$ (point group $O_h$) to the tetragonal $I4/mmm$ (point group $D_{4h}$) structure. In the absence of strain, our FM+SOC calculations with the Ni spin-polarization along the [001] direction (see Fig.~\ref{fig1}a), as is relevant to the measured AHC for the heterostructures, show a spin moment of 0.62 $\mu_B$/Ni atom in agreement with previous reports \cite{fu2019density}. The variation of the magnetic moment with strain is shown in Fig.~\ref{fig1}b.
Further, to understand the effect of strain on the AHE, we compute the AHC for bulk Ni both in the presence and absence of strain. The results of our calculations are also shown in Fig.~\ref{fig1}b. 
As seen from Fig.~\ref{fig1}b, the strain dependence of the computed AHC is qualitatively similar to that of the spin moment. Under compressive biaxial strain, both decrease monotonically with increasing strain magnitude, whereas under tensile strain, their magnitudes increase compared to the corresponding values in the pristine structure up to a certain strain value, beyond which both spin moment and AHC decrease.

Interestingly, our calculations show that the AHC is highest at $0.6\%$ strain, followed by $0.8\%$ and $0.3\%$ strain, corresponding to the Ni/STO, Ni/MgO, and Ni/LAO heterostructures, respectively (see Table~\ref{tab:table1}). This is in complete contrast to the measured value of AHC, shown in Fig.~\ref{fig_AHE}d. This discrepancy indicates that strain in fcc Ni alone is insufficient to capture the experimentally observed behavior, highlighting the importance of the substrate-induced symmetry-breaking interfacial effects in governing the AHE, as we proceed to discuss next.

\subsubsection{Interfacial Rashba Interaction}
To investigate the substrate-induced interfacial Rashba effects on the electronic structure and the resulting AHC, we now explicitly include the substrate in our simulations (see Section \ref{comp} for computational details) instead of bulk Ni, as discussed above.

\begin{figure}[t]
    \includegraphics[width=\columnwidth]{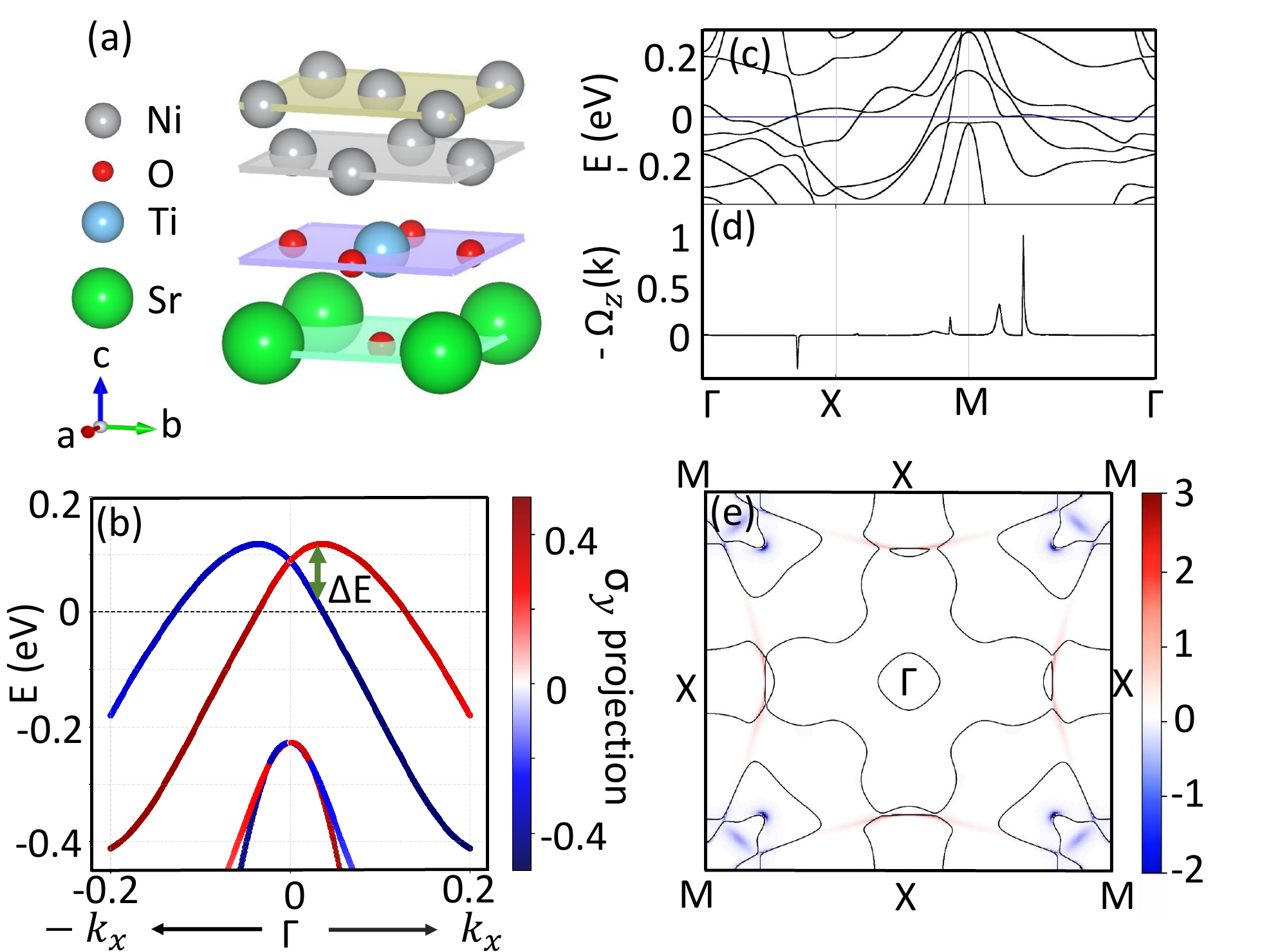}
    \caption{The Rashba interaction and the Berry curvature distribution in the Ni/STO interface. (a) The lowest energy configuration of the Ni/STO interface. (b) The non-magnetic spin-orbit-coupled band structure of Ni/STO. Here, the energy difference between the opposite spin-polarised bands, as indicated, is used to calculate the Rashba parameter. (c) The electronic band structure, and (d) the $k$-space distribution of the Berry curvature (in the unit of 10$^3$ \AA$^2$) in Ni/STO along the high-symmetry $k$ path, $\Gamma - X - M - \Gamma$. The Fermi energy in (b) and (c) is set at zero value. (e) The same momentum-space distribution of the Berry curvature (in units of 10$^3$ \AA$^2$) in Ni/STO on the $k_x - k_y$ plane.}
    \label{fig3}
\end{figure}

We begin our discussion with  Ni/STO as an example material. To isolate the substrate-driven effects, we construct the four possible interface structures by considering different terminating layers. All configurations are structurally optimized, and from the comparison of the total energies, we identify the structure shown in Fig.~\ref{fig3}a as the lowest energy interface structure (see SM~\cite{supple_mat} section 6). We note that the interfacial structure has $C_{4v}$ point group symmetry, indicating broken inversion symmetry. Furthermore, the presence of the TiO$_2$ terminating layer lowers the local symmetry of Ni atoms, making them inequivalent \cite{ohnishi2004preparation, kareev2008atomic}.

The broken inversion symmetry at the Ni/STO interface, in the presence of SOC, gives rise to the Rashba interaction of the form \cite{rashba1, rashba2},
\begin{equation}
H_{\mathrm{R}} = \alpha\, (\vec{\sigma} \times \vec{k}) \cdot \hat{z}.
\label{rashba_H}
\end{equation}
Here, $\alpha$, $\vec{\sigma}$, and $\vec{k}$ are respectively the Rashba parameter, Pauli spin matrices, and the crystal momentum. $\hat{z}$ corresponds to the direction perpendicular to the interface plane. 

The Rashba interaction leads to a characteristic linear-in-$k$ spin splitting in the electronic band structure with energy eigenvalues $\epsilon_{\pm} = \pm \alpha k$. 
This becomes evident from our computed band structure of the Ni/STO interface, as depicted in Fig.~\ref{fig3}b in the presence of SOC and in the absence of magnetization. For small values of $k$, the linear term dominates and the strength of the Rashba parameter $\alpha$ can be extracted from the energy separation $\Delta E$ between the spin-split bands, $\epsilon_+$ (red) and $\epsilon_-$ (blue), shown in Fig. ~\ref{fig3}b. For the Ni-$d$ bands (with quantum numbers, $j=5/2, m_j= \pm 2.5$) across the Fermi energy, our computed value of the Rashba parameter is $\alpha = 0.09$ eV\AA~, highlighting the presence of an intrinsic electric field at the interface due to broken inversion symmetry.

\begin{figure*}[htbp]
 \centering
    \includegraphics[width=2\columnwidth]{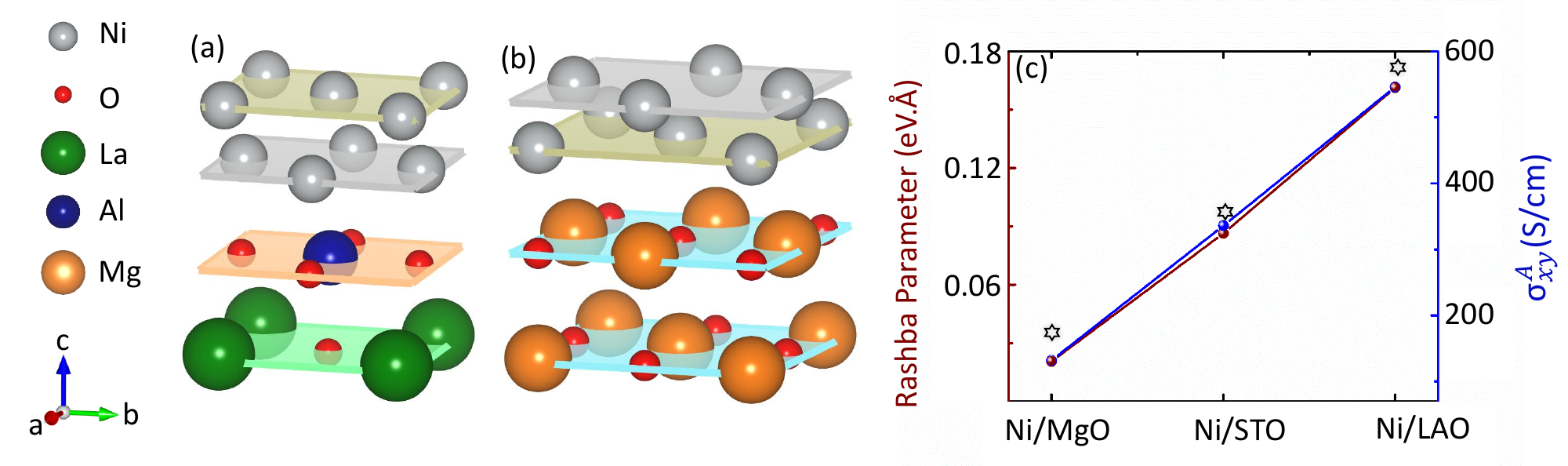}
    \caption{ Comparison of the Rashba parameter and the AHC for interfaces. Lowest energy configuration of (a) Ni/LAO, and (b) Ni/MgO. (c) The variation of Rashba parameter ($\alpha$) and AHC for Ni/MgO, Ni/STO, and Ni/LAO interfaces. Here, stars represent the experimentally obtained intrinsic AHC values for these interfaces.}
    \label{fig6}
\end{figure*}

We then introduce the effect of magnetism, and the resulting band structure of Ni/STO within FM+SOC is shown in Fig.~\ref{fig3}c. The broken time-reversal symmetry of the ferromagnetic Ni allows for a non-zero Berry curvature and, hence, AHC. 
We therefore compute these quantities using Wannier90, as detailed in section \ref{comp}. Our computed Berry curvature along a high-symmetry $k$ path, as well as the corresponding distribution on the $k_z=0$ plane, is shown in Fig.~\ref{fig3}d and e. We note that the Berry curvature distribution in Fig.~\ref{fig3}e has the four-fold rotational symmetry, consistent with the $C_{4v}$ point group symmetry of the Ni/STO heterostructure. 
The values of Berry curvature show the presence of a few sharp peaks at different $k$ points along the $M-\Gamma$ path of the Brillouin zone. These sharp peaks can be traced back to the SOC-avoided crossings near the Fermi energy, as shown in Fig.~\ref{fig3}c. 
The Brillouin zone sum of the Berry curvature gives the AHC.
Our computed value of AHC for Ni/STO interface is 336 S/cm, which is in reasonable agreement with the intrinsic AHC obtained from the measurements (see Fig.~\ref{fig6}c).

{\it Effect of substrate on the AHC-}To investigate if the interfacial effects are indeed important to understand the observed changes in AHC for different substrates, we explicitly carry out simulations for both Ni/MgO and Ni/LAO interfaces. Similarly to the case of Ni/STO, from the comparison of the total energies of all possible interface structures, depending on the terminating layers (see sections 7 and 8 of SM~\cite{supple_mat}), we identify the lowest energy structures of Ni/LAO and Ni/MgO interfaces, as shown in Fig.~\ref{fig6}a and b, respectively.

The lowest energy optimized structures of Ni/LAO and Ni/MgO interfaces have $C_{4v}$ symmetry, indicating the absence of inversion symmetry, similar to the Ni/STO interface. Since the broken inversion symmetry is key to the Rashba interaction, we compute the Rashba parameters for the same pair of Ni-$d$ bands with quantum numbers $(j=5/2, m_j= \pm 2.5)$ for Ni/LAO and Ni/MgO interfaces. The calculated values of the Rashba parameters are shown in Fig.~\ref{fig6}c for the three different choices of the substrate. Among these three, Ni/MgO has the weakest Rashba interaction, while the Ni/LAO interface has the strongest value. This can be understood from the fact that in Ni/MgO, the constituent elements are relatively light, which results in a weak SOC and hence a reduced magnitude of Rashba interaction. In contrast, Ni/LAO is a polar interface \cite{nakagawa2006some, popovic2008origin}, and our calculation shows that the interface undergoes a significant structural reconstruction to avoid polar catastrophe (see section 7 of SM~\cite{supple_mat}) \cite{pentcheva2009avoiding}. These interfacial polar distortions, together with the presence of a heavier atom, La, explain the computed larger Rashba interaction in Ni/LAO heterostructure.

We further compute the AHC for these interfaces, and the results of our calculations are shown in Fig.~\ref{fig6}c. We find that Ni/MgO exhibits the smallest AHC, while it is largest for Ni/LAO, following the same trend as the Rashba parameter ($\alpha$) across these interfaces. More interestingly, the computed AHC is also in reasonable agreement with the experimentally extracted intrinsic AHC values, as depicted in Fig.~\ref{fig6}c. These results clearly demonstrate that substrate-induced symmetry breaking manifests through the modifications of the Rashba interaction, which play a critical role in governing Hall transport properties in Ni heterostructures.

\subsection{Effect of Electric Field}

\begin{figure}[htbp]
 \centering
    \includegraphics[width=\columnwidth]{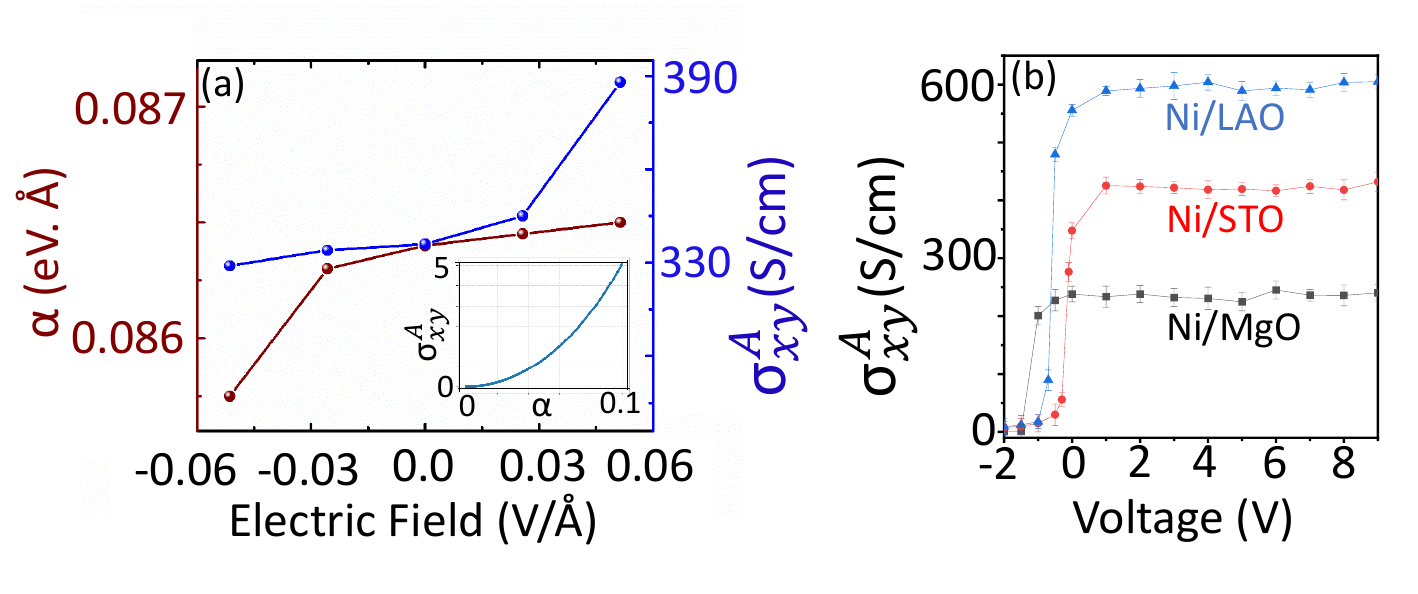}
    \caption{Electric field tuning of AHC. (a) The computed variation of the Rashba parameter ($\alpha$) and AHC with external electric field in the Ni/STO interface. Inset shows the variation of $\sigma_{xy}^A$ (in unit of 10$^{-3} \frac{e^2}{4\pi \hbar}$) as a function of $\alpha$ as given in Eq. (~\ref{two_ahc}). Here, the exchange splitting $\Delta$ is set to 1 eV. (b)  Experimental evolution of the maximum anomalous Hall conductivity $\sigma_{xy}^{A}$ with the gate voltage, demonstrating distinct electric-field tunability among the heterostructures.}
    
    \label{fig4}
\end{figure}

Since the AHC depends on the interfacial Rashba interaction, this suggests a possibility to tune the AHC by controlling the Rashba parameter using an external electric field. To illustrate this idea, we again consider Ni/STO as our example material. We investigate the system in the presence of an external electric field along the $\hat{z}$ direction with a magnitude ranging from -0.051 V/\AA~ to 0.051 V/\AA. Here, a negative value corresponds to a field applied along -$\hat{z}$, while a positive field is applied along the +$\hat{z}$. We optimize the crystal structure for each case to account for electrostatic screening.

To understand the effect of the external electric field on the Rashba interaction, we compute the band structure and extract the Rashba coupling coefficient $\alpha$ for the Ni-$d$ bands with quantum numbers $(j=5/2, m_j= \pm 2.5)$ as a function of the applied electric field. The results of our calculations are shown in Fig.~\ref{fig4}a. As evident from the figure, the magnitude of the Rashba coupling $\alpha$ can be controlled systematically by tuning the strength of the electric field.

The Rashba Hamiltonian in Eq. (~\ref{rashba_H}) can be rewritten in terms of a momentum-dependent magnetic field $\vec{B_k}$, $H_R = \vec{B}_k \cdot \vec{\sigma}$, where $\vec{B}_k = \alpha (\vec{k} \times \hat{z})$ \cite{z2}. Thus, a variation in $\alpha$ changes $\vec{B_k}$, and this change in the momentum-space magnetic field is anticipated to modify the AHC of the structure, as evident from the following low-energy two-band model,
\begin{equation}
\label{two_hamil}
\hat{H}
=
\frac{\hbar^2 k^2}{2m}\,\mathbb{I}
+
\alpha \, (\boldsymbol{\sigma} \times \mathbf{k}) \cdot \hat{\mathbf{z}}
-
\Delta \, \sigma_z.
\end{equation}
Here, the first term of the Hamiltonian denotes the parabolic bands of free electrons. The second term is the Rashba interaction due to the broken inversion symmetry, as introduced earlier, and the third term represents the exchange splitting due to the broken time-reversal symmetry of the ferromagnet. The low-energy model Hamiltonian captures the broken symmetries of the Ni/STO interface. The energy eigenvalues of the Hamiltonian (\ref{two_hamil}) are,~\(\varepsilon_{\pm}(\mathbf{k}) = \dfrac{\hbar^{2} k^{2}}{2m} \pm \sqrt{\alpha^{2} k^{2} + \Delta^{2}}\). Using these eigenvalues and the corresponding eigenvectors, we compute the $z$ component of the Berry curvature $\Omega^{z}(\mathbf{k})$ using the Kubo formula (see Eq. (~\ref{eq:berry}) in the computational details), and the result of our calculation is,
\begin{equation}
\label{two_berry}
\Omega^z_{\mp}(\mathbf{k})
=
\mp
\frac{\alpha^{2}\,\Delta}
{2\left(\alpha^{2}k^{2}+\Delta^{2}\right)^{3/2}}.
\end{equation}
Assuming that the Fermi energy lies within the exchange energy gap between two bands, viz., $-\Delta < E_F < \Delta $, and $ \Big(\frac{\alpha}{\Delta} \Big)^2 << 1$, we compute the AHC  by summing the Berry curvature over the occupied parts of the BZ (see Eq. (~\ref{eq:ahc}) of the computational details). The result is,
\begin{equation}
\label{two_ahc}
\sigma_{xy}
=
\frac{e^{2}}{4\pi\hbar}\,
\frac{\alpha^{2}}{\Delta}\,
\frac{\varepsilon_{F} + \Delta}
{\left(\frac{\Delta \hbar^{2}}{m} - \alpha^{2}\right)}.
\end{equation}
It is clear from the above expression that the AHC increases with increasing $\alpha$ (see the inset of Fig.~\ref{fig4}a), as anticipated earlier.  

To further confirm the changes in the AHC by tuning the Rashba parameter with an external electric field, we perform first-principles calculations of the AHC in the presence of different electric fields. The result of our calculations, as shown in Fig.~\ref{fig4}a, depicts a systematic variation in the AHC with the applied field, similar to the variation in Rashba coupling, discussed earlier. These results computationally establish the electric field as a tuning parameter for the AHC.

To experimentally probe this effect at room temperature, a bias electric field is applied across the substrate by connecting the top electrode to the Ni layer and the bottom electrode to the backside of the substrate. At the same time, a current is passed along the longitudinal direction of the film. The resulting longitudinal voltage (magnetoresistance) and transverse voltage (Hall signal) are measured simultaneously under the same applied electric bias field conditions. The bias voltage is systematically varied, including negative bias (down to  -3 V), where the electric field is directed from the film towards the substrate, and positive bias (up to 9 V), where the field is oriented towards the film. Furthermore, by sweeping the magnetic field, both the longitudinal resistivity ($\rho_{xx}$ vs H) and transverse (Hall) resistivity ($\rho_{xy}$ vs H) are obtained at different bias electric fields (see SM~\cite{supple_mat} section 9).

The application of an external electric field  markedly modulates the transverse Hall resistivity and longitudinal magnetoresistance of Ni/MgO, Ni/STO, and Ni/LAO heterostructures, as shown in section 9 of the SM~\cite{supple_mat,Feng2020, Mizuno2017, Kan2020}. The voltage-dependent maximum AHC value is again extracted from the Hall and magnetoresistance measurements using the same formula $\sigma_{xy} = -\frac{\rho_{xy}}{\rho_{xy}^2 + \rho_{xx}^2}$ and is also shown in Fig. \ref{fig4}b. On the positive-bias side, the saturation AHC follows the sequence Ni/LAO $>$ Ni/STO $>$ Ni/MgO. In contrast, under negative bias, the conductivity decreases, and the characteristic onset voltages across the samples are $\sim$ - 1.5 V for Ni/MgO, $\sim$ - 0.5 V for Ni/STO, and $\sim$ - 1 V for Ni/LAO, suggesting that this regime is governed primarily by the dielectric response of the substrates rather than the strain. 

The magnitude of this modulation is strongly influenced by dielectric screening of the substrate. For example, STO, with a very large dielectric constant ($\sim$300), produces a substantially stronger interfacial electrostatic perturbation than LAO ($\sim$25) or MgO ($\sim$9), thereby generating distinct boundary conditions for charge redistribution and transport tuning in each heterostructure.
Materials with larger dielectric permittivity exhibit stronger polarization under an applied field, which promotes more efficient interfacial charge accumulation and accelerates the transition toward a low-resistance state. Such rapid charge redistribution enhances the interfacial electric field and hence the Rashba effect~\cite{Bocirnea2020, Ying2017, Hyun2020, Hwang2012}. Taken together, these results demonstrate the electric-field-tunable spin-orbit-driven anomalous Hall transport properties.

\section{Summary and Outlook}

To summarize, using a combined theory and experimental approach, we have demonstrated that the AHC at Ni/oxide interfaces can be systematically tuned by modifying the Rashba interaction. Although biaxial strain is inherently introduced during the epitaxial growth of Ni thin films on MgO, STO, and LAO substrates, our first-principles DFT calculations reveal that the observed variations in AHC are primarily governed by interfacial electric fields rather than the lattice strain. The intrinsic electric fields directly influence the interfacial Rashba interaction, thereby modifying the Berry curvature of the electronic bands. Our DFT results, as well as the analytical modeling, establish that AHC is explicitly dependent on the Rashba parameter. This theoretical insight is further supported by experimental results, which show that applying an external electric field acts as an additional and reversible control knob for the AHC. 

Our work demonstrates that the magnitude and direction of the applied electric field strongly influence the Rashba interaction, which, in turn, modifies the AHC. These findings highlight interfacial Rashba engineering as a key mechanism for controlling Berry curvature-driven transport phenomena. This methodology can be readily applied to a wider range of transition-metal/oxide interfaces, enabling electrically tunable Hall responses independent of magnetic field control. Furthermore, the demonstrated electrical tunability of the AHE suggests the possibility of employing ferroelectric substrates, where the switchable electric polarization could provide a reversible control of the Rashba interaction \cite{Fabian2024}. Investigating such Ni/ferroelectric oxide heterostructures is therefore an interesting direction for future studies. From a device perspective, our results offer a feasible design principle for next-generation spintronic devices based on controllable anomalous Hall phenomena.

\section{METHODOLOGY}
\subsection{Experimental Methods}
High-purity (99.99$\%$) commercial target of Nickel (Ni) (Kurt J. Lesker) is used for thin film growth. A single-layer Ni is fabricated on commercial single-crystal (MTI Corporation) MgO (001), STO (001), and LAO (001) substrates using DC magnetron sputtering (Minilab S80A, Moorfield Nanotechnology). During the deposition, the base pressure was $\sim 5 \times 10^{-7}$ mbar, with an Ar working pressure $\sim$ 0.06 mbar and input power $\sim$ 35 W. The substrates are rotated and maintained at 450$^{\circ}$C to ensure uniform growth and crystallization. The Ni layer thickness is fixed at 35 nm with a deposition rate of $\sim$ 0.3 \AA/s. After deposition, films are annealed at 450$^{\circ}$C for 1 h under high vacuum ($\sim$ 10$^{-7}$ mbar) and then cooled to room temperature.\\
The epitaxial growth and crystalline structure of the films are examined using reflection high-energy electron diffraction (RHEED; STAIB Instruments, Germany, operated at 30 keV, and analyzed using KSA400 software) and X-ray diffraction (XRD; Rigaku Smart Lab, Cu K$_\alpha$, $\lambda$ = 1.5406 \AA, five-axis goniometer). Magnetic and transport properties are measured using a Physical Property Measurement System (PPMS, Quantum Design DynaCool 9T), including M(H) loops and Hall resistivity. Diamagnetic substrate contributions are subtracted from all magnetization data.\\

\subsection{Computational Details}\label{comp}
 
The electronic and magnetic properties of the bulk Ni both in the absence and presence of strain, and Ni/STO, Ni/MgO and Ni/LAO interfaces in the presence and absence of electric field are computed using the QUANTUM ESPRESSO code~\cite{giannozzi2009quantum, giannozzi2017advanced} within the local density approximation ~\cite{parr1989density}, employing norm-conserving pseudopotentials. A kinetic energy cutoff of 150 Ry and a Monkhorst-pack k-grid of $12 \times 12 \times 12$ are chosen for the plane wave basis set in bulk Ni case (both with and without strain) to achieve convergence. On the other hand, an energy cutoff of 160 Ry and a $12 \times 12 \times 2$ $k$ mesh are used for Ni/MgO, Ni/STO, and Ni/LAO interfaces. The width of the Gaussian broadening parameter is set to 0.01 Ry in all calculations to account for the partial occupancy of the bands near the Fermi energy.

For the calculations of bulk Ni in the presence of strain, we compute the strain value (in $\%$) as $(a-a_0)/a_0 \times 100$, where $a_0$ is the bulk Ni lattice parameter and $a$ is the in-plane lattice parameter in the presence of biaxial strain. In bulk Ni, the structural relaxation has been performed until the Hellmann-Feynman forces on each atom reduce below $5 \times 10^{-4}$ Ry/\AA. For the structural relaxation of bulk Ni in the presence of biaxial strain, we also relax the out-of-plane lattice constant, allowing for changes in volume.

To model the minimal heterostructure, we consider a single unit cell of Ni on top of a single unit cell of STO, MgO, and LAO (see SM~\cite{supple_mat} for details). An empty space of 20 \AA~is used along the out-of-plane direction to avoid spurious interactions between periodic images. Since the periodic images may introduce artificial electrostatic potential, the dipole correction has been included in the calculations. After constructing the Ni/MgO, Ni/STO, and Ni/LAO interfaces, the structural relaxation is performed with Hellmann-Feynman force threshold of $5 \times 10^{-4}$ Ry/\AA~keeping the in-plane lattice parameters fixed. In the case of Ni/STO, an electric field is applied to the interface along the out-of-plane direction (z-axis) via a sawtooth potential with the change in slope, located far from the interface. In the presence of extrinsic electric field, the structural relaxation is performed for Ni/STO using the same methodology as mentioned above.

To calculate the Berry curvature and AHC, we use the Wannier90 \cite{pizzi2020wannier90} code with a properly chosen basis set for the respective systems to obtain maximally localized Wannier functions (MLWFs). For bulk Ni both in the absence and presence of strain, Ni $d$ and $s$ orbitals are included in the Wannierization procedure. To construct the MLWFs for Ni/MgO, we chose $p_z$ orbitals of Mg, $s$ and $p$ orbitals of O and $d$ orbitals of Ni. In the Wannierization process of Ni/LAO, the $d_{x^2-y^2}$ and $d_{z^2}$ orbitals of La, $s$ and $p$ orbitals of O, and Ni $d$ orbitals are chosen. In Ni/STO, the MLWFs are constructed from Ni $d$ and $s$, O $s$ and $p$, and Ti $d_{xy}$, $d_{yz}$, $d_{xz}$ orbitals. The same orbitals are used for constructing the MLWFs in Ni/STO in the presence of an external electric field. To mimic the experimental set up for the AHC in the presence of a magnetic field along $\hat z$, under which the Ni spin moments reorient to the out-of-plane direction, we consider the Ni spins to be directed along $\hat z$ for the calculation of AHC. We note that, however, in the absence of the external magnetic field, the Ni spin moments prefer to orient in the in-plane direction (see SM~\cite{supple_mat} for details).

Subsequently, we compute the AHC of these structures by summing the Berry curvature over the occupied part of the Brillouin zone \cite{becur1, becur2}, 
\begin{equation}
\label{eq:ahc}
\sigma_{xy}^{\mathrm{AHC}}
=
-\frac{e^{2}}{\hbar}
\frac{1}{N_k V_c}
\sum_{n,\vec{k}}
\Omega^{z}_{n}(\vec{k}).
\end{equation}
Here, $V_c$ is the unit cell volume and $N_k$ denotes the total number of
$\vec{k}$-points used in the Brillouin-zone sampling. The summation is considered over all electronic states below the Fermi energy. We check the convergence of AHC by varying the number of $k$ points. We achieve the convergence in the AHC with a $240\times240\times240$ $k$ mesh for bulk Ni both in the absence and presence of strain, while a $240\times240\times40$ $k$ mesh is used to obtain a converged AHC in Ni/MgO, Ni/LAO, and Ni/STO (both with and without the electric field). An adaptively refined mesh of $7 \times 7\times 7$ is used around $k$ points with Berry curvature value exceeding $100$~\AA$^2$.

The Berry curvature $\Omega^{z}_{n}(\vec{k})$ of the $n$-th band can be computed using the Kubo formula,
\begin{equation}
\label{eq:berry}
\Omega^{z}_{n}(\vec{k})
=
-2\hbar^{2}
\sum_{n' \neq n}
\frac{
\mathrm{Im}
\left[
\langle \psi_{n\vec{k}} | v_x | \psi_{n'\vec{k}} \rangle
\langle \psi_{n'\vec{k}} | v_y | \psi_{n\vec{k}} \rangle
\right]
}{
\left( \epsilon_{n'\vec{k}} - \epsilon_{n\vec{k}} \right)^2
}.
\end{equation}

Here, the velocity operator is expressed as
\begin{equation}
v_\alpha
=
\frac{1}{\hbar}
\frac{\partial H}{\partial k_\alpha},
\qquad \alpha = x, y, z.
\end{equation}
$\vec{k} = (k_x, k_y, k_z)$ denotes the crystal momentum. $H$, $\epsilon_{n\vec{k}}$, and $\psi_{n\vec{k}}$ are respectively the Hamiltonian, and the corresponding energy eigenvalues, and eigenstates.

\section*{acknowledgments}
D.M. and S.B. thank the Anusandhan National Research Foundation, Government of India (Grant No. ARNF/ARG/2025/007161/PS). M.G. and S.B. thank National Supercomputing Mission for providing computing resources of ‘PARAM Rudra’ at IIT Bombay, implemented by CDAC and supported by the Ministry of Electronics and Information Technology (MeitY) and Department of Science and Technology, Government of India. S.B. gratefully acknowledges financial support from the IRCC Seed Grant (Project Code: RD/0523-IRCCSH0-018), the INSPIRE Research Grant (Grant No.- DST/INSPIRE/IFF/BATCH-20/2024-25/IFA 23-PH 299), and the ANRF PMECRG Grant (Grant No.- ANRF/ECRG/2024/001433/PMS). D.M.  acknowledges financial support from the Technical Research Center, Department of Science and Technology (DST), Government of India (Grant No. AI/1/62/IACS/2015), as well as from the India–Russia Joint Research Call, DST, Government of India (Grant No. DST/IC/RSF/2024/542).

\bibliographystyle{apsrev4-2}
\bibliography{a1}

\end{document}